\documentclass[preprint,12pt]{elsarticle}

\usepackage{amssymb}
\usepackage{enumerate}
\usepackage{array}
\usepackage{amsthm}
\usepackage{amsmath}
\usepackage{units}
\usepackage{color}
\usepackage{textcomp}
\usepackage{minted}
\usepackage{tcolorbox}
\usepackage{xpatch}

\usepackage{booktabs}

\usepackage[colorlinks=false,hidelinks]{hyperref}

\usepackage{subcaption}

\makeatletter
\def\dontdofcolorbox{\renewcommand\fcolorbox[4][]{##4}}
\xpatchcmd{\inputminted}{\minted@fvset}{\minted@fvset\dontdofcolorbox}{}{}
\makeatother

\biboptions{sort&compress}

\newcounter{bla}

\journal{Computer Physics Communications}
\begin{document}

\begin{frontmatter}



%\title{MCPlas -- a toolbox for non-thermal plasma modelling using COMSOL Multiphysics}
%\title{MCPlas for plasma modelling using (reusable / standardized / ... ) input schemas with COMSOL Multiphysics}

\title{MCPlas, a MATLAB toolbox for reproducible plasma modelling with COMSOL}


\author[1]{Marjan N. Stankov}\corref{cor1}
\author[2]{Daan Boer}
\author[3]{Wouter Graef}
\author[3]{Kevin van ‘t Veer}
\author[1]{Aleksandar P. Jovanovi\'{c}}
\author[1]{Florian Sigeneger}
\author[1]{Detlef Loffhagen}
\author[2]{Jan van Dijk}
\author[1]{Markus M. Becker}

\cortext[cor1] {Corresponding author.\\\textit{E-mail address:} marjan.stankov@inp-greifswald.de}

\affiliation[1]{organization={Leibniz Institute for Plasma Science and Technology (INP)},%Department and Organization
            addressline={Felix-Hausdorff-Str. 2}, 
            city={Greifswald},
            postcode={17489}, 
            %state={},
            country={Germany}}
\affiliation[2]{organization={Eindhoven University of Technology},%Department and Organization
            addressline={PO Box 513}, 
            postcode={5600 MB}, 
            city={Eindhoven},
            %state={},
            country={Netherlands}}
\affiliation[3]{organization={Plasma Matters B.V., Campus Eindhoven University of Technology, DIFFER building},%Department and Organization
            addressline={De Zaale 20}, 
            postcode={5612 AJ}, 
            city={Eindhoven},
            %state={},
            country={Netherlands}}

\begin{abstract}

The MCPlas toolbox represents a collection of MATLAB functions for the automated generation of an equation-based fluid-Poisson model for non-thermal plasmas in the multiphysics simulation software COMSOL. 
Following the development of the new generation of the LXCat platform, all input data are prepared in a structured and interoperable JSON format and can be supplied and validated using existing JSON schemas.
The toolbox includes fully transparent, editable MATLAB source code and offers an advanced description of electron transport in addition to commonly used approaches in the plasma modelling community. 
It supports one-dimensional and two-dimensional modelling geometries employing Cartesian, polar and cylindrical coordinate systems. 
MCPlas is tested on two reference cases: DC- and RF-driven low-pressure glow discharges in argon. Comparison of MCPlas results with results obtained by employing COMSOL's Plasma Module verifies the reliability of the plasma model implemented by MCPlas and demonstrates the significance of electron transport treatment and boundary conditions applied in the toolbox.
Using the same examples, the easy handling of complex reaction kinetic models in MCPlas and the reusability of its JSON input data across different modelling platforms are illustrated. This demonstrates that MCPlas provides a transparent and reproducible workflow for the simulation of non-thermal plasmas using COMSOL.
%Using the same examples, it is shown that MCPlas provides a transparent and reproducible workflow for setting up individual plasma models in COMSOL.

\end{abstract}


\end{frontmatter}

% \linenumbers

% All CPiP articles must contain the following
% PROGRAM SUMMARY.

%{\bf PROGRAM SUMMARY/NEW VERSION PROGRAM SUMMARY}
  %Delete as appropriate.
{\bf PROGRAM SUMMARY}

\begin{small}
\noindent
{\em Program title: MCPlas}                                          \\
{\em CPC Library link to program files:} (to be added by Technical Editor) \\
{\em Developer's repository link:} (https://github.com/INP-SDT/MCPlas) \\
{\em Code Ocean capsule:} (to be added by Technical Editor)\\
{\em Licensing provisions:} MIT  \\
{\em Programming language:} MATLAB \\
%{\em Supplementary material:} \\
{\em Nature of problem: }\\
 Fluid-Poisson simulations of low-temperature plasmas are widely used but often lack reproducibility, transparency and interoperability. In most commercial and in-house tools, the governing equations, boundary conditions and reaction kinetics are hidden in proprietary project files or ad hoc scripts, and complex reaction kinetics models (RKMs) must be entered manually. Moreover, there is no machine-readable community standard or format for plasma chemistries and related input data, so the same RKM cannot be reliably reused across different modelling platforms. As a result, the precise reconstruction and verification of published models is difficult, comparison between tools is hindered, and the same chemistry cannot be reliably reused across different codes, conflicting with the FAIR (Findable, Accessible, Interoperable, Reusable) data principles. \\
{\em Solution method:}\\
MCPlas addresses these issues by standardising all input data and automating the implementation of a fluid-Poisson model for non-thermal plasmas in COMSOL. All model-defining information, such as species states and properties, RKM, transport and rate coefficients, model setup and numerical parameters, is stored in structured JSON documents. These are validated against JSON schemas aligned with the new LXCat data model and the plasma metadata schema Plasma‑MDS. From these schema-validated JSON files, MCPlas automatically builds fully equation-based fluid-Poisson models in COMSOL via the LiveLink\textsuperscript{\texttrademark} \textit{for} MATLAB\textsuperscript{\fontsize{4}{4}\selectfont\textregistered} module, with explicit access to the implemented equations and boundary conditions. The toolbox offers options for advanced electron transport description and boundary treatments in addition to commonly used formulations. The same JSON-based model definitions can be used unchanged in other software packages for plasma modelling, enabling interoperable, cross-platform simulations and implementing a transparent, FAIR-compliant workflow for reproducible plasma modelling. \\
 {\em Additional comments including restrictions and unusual features:} The commercial software packages MATLAB\textsuperscript{\fontsize{4}{4}\selectfont\textregistered} (R2023a), COMSOL Multiphysics\textsuperscript{\fontsize{4}{4}\selectfont\textregistered} (v.6.2) and the COMSOL module LiveLink\textsuperscript{\texttrademark} \textit{for} MATLAB\textsuperscript{\fontsize{4}{4}\selectfont\textregistered} were used for development MCPlas and testing generated models. There should be no restrictions on using MCPlas with other software versions.\\
  %Provide any additional comments here.
   \\

%\begin{thebibliography}{0}
%\bibitem{1}Reference 1         % This list should only contain those items referenced in the                 
%\bibitem{2}Reference 2         % Program Summary section.   
%\bibitem{3}Reference 3         % Type references in text as [1], [2], etc.
                               % This list is different from the bibliography at the end of 
                               % the Long Write-Up.
%\end{thebibliography}
%* Items marked with an asterisk are only required for new versions of programs previously published in the CPC Program Library.\\
\end{small}

\section{Introduction}
\label{sec:Intr}

In the last decades, fluid modelling has been established as a standard tool for the theoretical analysis of low-temperature plasmas (LTP)~\cite{Colonna_book}.  
%Fluid (or hydrodynamic) models are valid if the  mean free path of the different species in the plasma is much smaller than the characteristic dimension of the discharge~\cite{Belenguer_1990}. 
State-of-the-art fluid models for non-thermal LTP include balance equations for the number densities of particles and mean electron energy, hydrodynamic transport equations for heavy particles and the Poisson equation for the determination of the electric field~\cite{Alves_2012, Becker-2013-J-Phys-D, Markosyan_2015, Alves_2018,Stankov_2020, Wang_2021,Stankov_2022, Levko_2023}.
%Here, an accurate description of electron energy transport is crucial for the reliability of any fluid model~\cite{Becker_2013_AIP}. 
The gain and loss of charge carriers and neutral particle species is usually described on the basis of rate equations considering the relevant collision and radiation processes for the plasma species~\cite{Bogaerts_1998}. 
Due to the wide range of applications and popularity of fluid models, many different methods for solving their equations have been developed. Nowadays, they are often a part of various commercial software packages, e.g., PLASIMO~\cite{plasimo,vanDijk-2009-ID2562}, COMSOL Multiphysics\textsuperscript{\fontsize{4}{4}\selectfont\textregistered} (in further text COMSOL)~\cite{Comsol}, CFD-ACE~\cite{CFD-ACE}, and OverViz~\cite{OverViz}. 
In particular, the COMSOL's Plasma Module (CPM) is very popular in the low-temperature plasma physics community~\cite{Rafatov_2012, Hong_2017, Baeva_2018, Murakami_2020, Terentev_2022, Bose_2022, Datta_2024}. 
The main advantages of these commercially available software packages are the relatively fast implementation of the plasma model even for complex geometries and the access to efficient numerical solvers. 
Furthermore, some packages provide the possibility to automatically include a certain set of reaction processes and direct access to physically motivated boundary conditions~\cite{Comsol}. 
Such features make complex plasma models accessible even to users without much experience in plasma modelling, providing them the opportunity to take the benefits of established modelling tools relatively easy. 
The disadvantage of using commercial software packages for plasma modelling is that, in most cases, the basic equations and boundary condition used to describe the plasma properties cannot freely be changed. 
For example, recently published models including an alternative description of electron~\cite{Becker-2017-ID4159,Stankov_2022} or ion~\cite{Semenov_2017} transport are not readily available if commercial software packages are used. 

Regardless of the choice of the software, modelling of non-thermal plasmas requires defining the properties of the considered species and reaction kinetic processes in terms of a reaction kinetic model (RKM).
Especially when dealing with complex plasma chemistries considering a large number of particle species and reactions, this becomes a time-consuming and error-prone process.
Moreover, the manual implementation of species properties and their reactions in the software system complicates the reproduction and verification of modelling results. 
The reliance on manual input does not only increase the potential for inconsistencies and errors, but also hampers the ability to replicate findings and verify the accuracy of the model, thereby undermining the reliability of the results produced.

These challenges are addressed by the open-source MATLAB-COMSOL toolbox for plasma modelling (MCPlas) presented here.
MCPlas was developed using the MATLAB\textsuperscript{\fontsize{4}{4}\selectfont\textregistered}~\cite{Matlab} (in further text MATLAB) programming language for the automated build-up of an equation-based fluid-Poisson model in COMSOL. For that purpose, it uses the linking between MATLAB and COMSOL provided by the COMSOL LiveLink\textsuperscript{\texttrademark} \textit{for} MATLAB\textsuperscript{\fontsize{4}{4}\selectfont\textregistered} module~\cite{Jovanovic_2021}. 
It aims at an easy and fast generation of models for the spatio-temporal modelling of non-thermal plasmas based on the input data containing all information about the considered species and RKM.
Earlier (closed-source) versions of MCPlas have been used for the simulation of streamers and dielectric barrier discharges at atmospheric pressure~\cite{Bagheri-2018-ID5240,Jovanovic_2021}.
It has been shown in~\cite{Jovanovic_2021} that a notable advantage of MCPlas over many commercial software packages lies in the simplified model setup as well as the transparent and direct access to all MATLAB source files, i.e.\ the specific model equations, boundary conditions and all input data.
This accessibility empowers users to tailor the implemented model to meet their specific needs without losing the reproducibility and transparency of the model and the results.

From its first deployment, the focus of the further development of
MCPlas has been set on the standardization of the input files.
%As the MCPlas toolbox requires input data related to RKM, 
An important point that arose during the development was the selection of the input data format for the RKM. 
In the LTP community, there is no specific file format for preparing RKM data and a standardised process for organising them~\cite{Alves_2023}. 
Usually, each scientific group uses different modelling tools and input file types, which complicates sharing and the reuse of data. 
However, the LXCat open-access website~\cite{LXcat_paper_2021}, renowned for storing and exchanging data essential for modelling of LTP, has proposed JSON (JavaScript Object Notation)~\cite{JSON_website, JSON_book} as the exchange format in its latest platform version, currently undergoing active, open-source development~\cite{lxcat-zenodo,lxcat_github}.
JSON is a well-established and widely adopted data-interchange format in computer science. 
It is platform-independent and readable for humans and machines.
One of the key benefits of the JSON data format is the ability to easily validate the structure and properties of the data using a JSON schema. 
Considering the strong connection between the RKM input data required for the MCPlas toolbox and the data available on the LXCat platform, the JSON data format stood out as the most fitting choice for defining input data for MCPlas.
Therefore, the published open-source version of MCPlas presented here follows the LXCat developments. 
This supports the adoption of the FAIR (Findable, Accessible, Interoperable, Reusable) data principles~\cite{Wilkinson-2016-ID142} in LTP research as well as the reproducibility of modelling results as demonstrated by several use cases in the present work.

The manuscript is structured as follows: the subsequent section provides a comprehensive description of the MCPlas toolbox, encompassing the details about the modelling approach, the architecture of MCPlas, and detailed instructions for using the toolbox.
Within the same section, all details about the input data are included, covering the organization and management of JSON files containing information on the RKM, properties and transport coefficients of all particle species and some general input information.  
In section \ref{sec:results}, a comparison of modelling results obtained from the COMSOL model generated by the MCPlas toolbox and the CPM is shown with the examples of direct-current (DC) and radio-frequency (RF) glow discharges in argon.
This includes the code verification by direct comparison of results as well as the demonstration of beneficial features of the modelling approach applied in MCPlas and simple handling of complex RKMs.
In addition, the reusability of the RKM input data in the JSON format is demonstrated by using the same input files in two other plasma simulation tools and repeating the simulation of the same examples.
Finally, conclusions are drawn in section~\ref{sec:conclusions}.

\section{MCPlas toolbox description}
\label{MCPlas_toolbox_description}

The MCPlas toolbox has been developed for the equation-based implementation of fluid-Poisson models in COMSOL.
It consists of a set of MATLAB scripts and functions and uses the COMSOL LiveLink\textsuperscript{\texttrademark} \textit{for} MATLAB\textsuperscript{\fontsize{4}{4}\selectfont\textregistered} module to set up a COMSOL file containing the entire model and all input data. With this, the generated models are ready to be used with the General Equation Module as a part of the basic COMSOL package without the need for CPM or any other COMSOL modules. Note that for the development of the MCPlas toolbox, MATLAB version R2023a and COMSOL version 6.2 were used. Both older and newer versions of the software were tested for the purpose of using MCPlas and no problems were encountered, so there should be no restrictions on using MCPlas with other versions.

The input data necessary for running MCPlas includes information about the RKM, transport properties of the particle species, as well as general information about the plasma source, plasma medium,
and fluid-Poisson model.
These data represent a heterogeneous data set that requires a standardised schema to be verified. 
Here, the ability of the JSON data format to store objects with different sets of attributes, which can be validated against a schema document, is used to define and organise all input data, similar to the approach suggested in~\cite[Section~6.1]{LXcat_paper_2021}.

\subsection{Modelling geometries}
\label{modelling_geometries}
\begin{figure}
	\centering
	\includegraphics[]{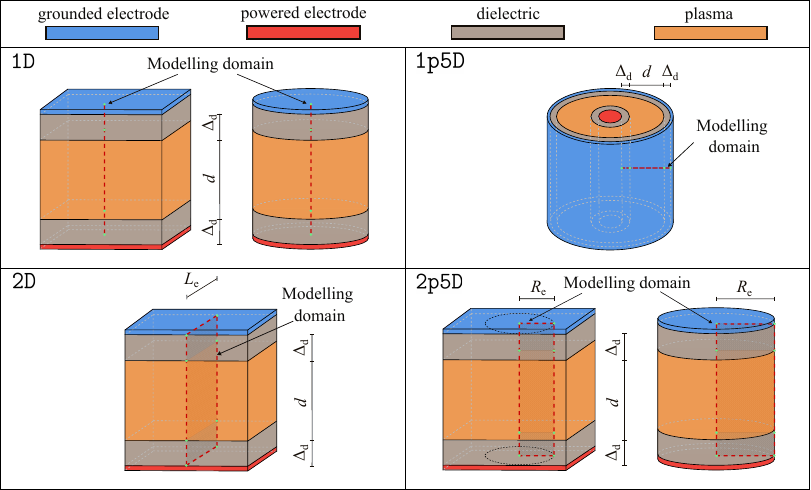}
	\caption{Modelling geometries supported by MCPlas toolbox ($d$ - discharge gap, $\Delta_\mathrm{d}$ - dielectric thickness, $L_\mathrm{e}$ - electrode length and $R_\mathrm{e}$ - electrode radius).}
	\label{fig:figure_1}
\end{figure}

MCPlas enables the generation of models applicable for time-dependent, spatially one- and two-dimensional modelling of non-thermal plasmas driven by an electric field between two metal electrodes, with optional dielectric layers on their surfaces. 
Users can choose among four modelling geometries illustrated in figure~\ref{fig:figure_1}, so far. 
The \texttt{1D} or \texttt{1p5D} option,  which consider Cartesian or polar coordinates, respectively, can be used for generation of one-dimensional models.
For two-dimensional models, the \texttt{2D} option, utilising Cartesian coordinates, and the \texttt{2p5D} option, employing cylindrical coordinates, are available. 
The provided modelling geometries are well suited to plane-parallel and coaxial plasma sources. However, they can also be applied to other configurations according to user's needs.

\subsection{Basic equations}
\label{basic_equations}

\subsubsection{Fluid-Poisson model}
\label{fluid_poisson_model}

The plasma description provided by MCPlas is based on the common fluid-Poisson model~\cite{Lymberopoulos_1993}
\begin{eqnarray}
    \frac{\partial}{\partial t}n_j
        + \nabla\cdot\mathbf{\Gamma}_j = S_j,
        \label{eq:continuity}\\
    \frac{\partial}{\partial t}w_\mathrm{e} 
        +\nabla\cdot\mathbf{Q}_\mathrm{e}
        = -e_0\mathbf{\Gamma}_\mathrm{e}\cdot \mathbf{E} + \tilde{S}_\mathrm{e},
        \label{eq:we}\\
    -\nabla \cdot(\varepsilon_\mathrm{r}\varepsilon_0\nabla\phi)
        = \sum_j q_j n_j,
        \label{eq:poisson}
\end{eqnarray}
where~\eqref{eq:continuity} represents the balance equations for the particle number densities $n_j$ of species with index $j$ (electrons, ions, neutrals),
%(electrons and neutral and charged heavy particles)
charge $q_j$ and particle flux $\mathbf{\Gamma}_j$,~\eqref{eq:we} is the balance equation for the energy density $w_\mathrm{e}=n_\mathrm{e}u_\mathrm{e}$ of electrons ($j = \mathrm{e}$) with the mean electron energy $u_\mathrm{e}$ and energy flux $\mathbf{Q}_\mathrm{e}$, and~\eqref{eq:poisson} represents the Poisson equation for the self-consistent determination of the electric potential $\phi$ and electric field $\mathbf{E}=-\nabla\phi$. 
Here, $e_0$, $\varepsilon_\mathrm{r}$ and $\varepsilon_0$ are
the elementary charge, the relative permittivity of the medium and the vacuum permittivity, respectively.
The source terms $S_j$ describe the gain and loss of particles due to collision and radiation processes, and $\tilde{S}_\mathrm{e}$ accounts for the corresponding gain and loss of electron energy. 
It should be emphasized that all variables ($n_j$, $w_\mathrm{e}$, $\mathbf{\Gamma}_j$, $\mathbf{Q}_\mathrm{e}$, $S_j$,  $\tilde{S}_\mathrm{e}$ and $\mathbf{E}$) are space- and time-dependent quantities. To improve clarity and readability, the explicit notation of their dependence on space coordinate $\mathbf{r}$ and time $t$ is suppressed in the text. 

\subsubsection{Particle and electron energy fluxes}
\label{particle_fluxes}

The fluxes ${\mathbf{\Gamma}}_\mathrm{h}$ of heavy particles ($j=\mathrm{h}$) in equation~\eqref{eq:continuity} are expressed by the common drift-diffusion approximation%~\cite{Sigeneger_1999, Grubert-2009-ID2551}
\begin{equation}
\mathbf{\Gamma}_\mathrm{h}
= \mathrm{sgn}(q_\mathrm{h})n_\mathrm{h} b_\mathrm{h} \mathbf{E}   
-D_\mathrm{h}\nabla n_\mathrm{h}\,, \label{eq:flux_particle}
\end{equation} 
where,  $b_\mathrm{h}$ and $D_\mathrm{h}$ stand for the  mobility and diffusion coefficient of heavy species $\mathrm{h}$, respectively, while the function $\mathrm{sgn}(q_\mathrm{h})$ defines the sign of $q_\mathrm{h}$.
Three options are offered by MCPlas for the definition of the electron flux ${\mathbf{\Gamma}}_\mathrm{e}$ and electron energy flux ${\mathbf{Q}}_\mathrm{e}$. 
The conventional drift-diffusion approximation (option \texttt{DDAc}) for these fluxes reads~\cite{Sigeneger_1999, Grubert-2009-ID2551}
\begin{eqnarray}
\mathbf{\Gamma}_\mathrm{e}
= -n_\mathrm{e} b_\mathrm{e} \mathbf{E}   
-\nabla(D_\mathrm{e}n_\mathrm{e})\,,\label{eq:JeDDAc}\\
\mathbf{Q}_\mathrm{e}
= -w_\mathrm{e} \tilde{b}_\mathrm{e} \mathbf{E}   
-\nabla(\tilde{D}_\mathrm{e}w_\mathrm{e})\,,\label{eq:QeDDAc}
\end{eqnarray} 
where $b_\mathrm{e}$ and $D_\mathrm{e}$ are electron transport coefficients, and $\tilde{b}_\mathrm{e}$ and $\tilde{D}_\mathrm{e}$ represent the electron energy transport coefficients. The commonly used approach (option \texttt{DDA53}) employs (\ref{eq:JeDDAc}) together with the simplified form of the electron energy flux

\begin{equation}
\mathbf{Q}_\mathrm{e}
= -\frac{5}{3} w_\mathrm{e} {b}_\mathrm{e} \mathbf{E}   
-\frac{5}{3}\nabla({D}_\mathrm{e}w_\mathrm{e})\,.\label{eq:QeDDA53}
\end{equation}

The novel drift-diffusion approximation (option \texttt{DDAn}) represents a third way of specifying ${\mathbf{\Gamma}}_\mathrm{e}$ and  ${\mathbf{Q}}_\mathrm{e}$. It has been deduced by an expansion of the electron velocity distribution function (EVDF) in Legendre polynomials and the derivation of the first four moment equations from the electron Boltzmann equation~\cite{Becker-2017-ID4159, Becker_2013_AIP,Becker_2013_APM}. It reads

\begin{eqnarray}
\mathbf{\Gamma}_\mathrm{e}
= -\frac{e_0}{m_\mathrm{e}\nu_\mathrm{e}}\nabla
\Bigl((\xi_0 + \xi_2)n_\mathrm{e} \Bigr)
-\frac{e_0}{m_\mathrm{e}\nu_\mathrm{e}} \mathbf{E} n_\mathrm{e}\,, \label{eq:JeDDAn}\\
\mathbf{Q}_\mathrm{e} = -\frac{e_0}{m_\mathrm{e}\tilde{\nu}_\mathrm{e}}\nabla
\Bigl((\tilde{\xi}_0 + \tilde{\xi}_2) w_\mathrm{e} \Bigr) \label{eq:QeDDAn}\\
\qquad\qquad\; -\frac{e_0}{m_\mathrm{e}\tilde{\nu}_\mathrm{e}}\Bigl(\frac{5}{3}
+ \frac{2}{3}\frac{\xi_2}{\xi_0}\Bigr)\mathbf{E} w_\mathrm{e}, \nonumber
\end{eqnarray}
and includes the momentum and energy flux dissipation frequencies $\nu_\mathrm{e}$ and $\tilde{\nu}_\mathrm{e}$, respectively, the transport coefficients $\xi_0$, $\xi_2$, $\tilde{\xi}_0$ and $\tilde{\xi}_2$, as well as the electron mass $m_\mathrm{e}$. 
It should be emphasized that this approximation is unique to the MCPlas toolbox, as to our knowledge it is not part of any other modelling tool.
Considering the accuracy improvements relative to the drift-diffusion approximation at low and atmospheric pressuress~\cite{Becker_2013_AIP,Baeva-2019-ID5345}, it represents a highly significant feature of the toolbox.
Details about the used transport coefficients in all presented approximations are given in the section~\ref{transport_and_rate_coefficients}.

\subsubsection{Boundary conditions}
\label{boundary_fluxes}

Boundary conditions for the balance equations for electron density (\ref{eq:continuity})  and electron energy density (\ref{eq:we}) are included in MCPlas in accordance with Hagelaar~\textit{et al.}~\cite{Hagelaar_2000} and read

\begin{eqnarray}
\mathbf{\Gamma}_\mathrm{e}\cdot\boldsymbol{\nu}
= \frac{1-r_\mathrm{e}}{1+r_\mathrm{e}}
\Bigl(\left|n_\mathrm{e} \mathbf{v}_{\mathrm{d},\mathrm{e}}\cdot\boldsymbol{\nu} \right|
+\frac{1}{2}n_\mathrm{e} v_{\mathrm{th},\mathrm{e}}\Bigr)
-\frac{2}{1+r_\mathrm{e}}\gamma\sum_i\max(\mathbf{\Gamma}_i\cdot\boldsymbol{\nu},0)
\label{eq:boundary_e}\,,\\
\mathbf{Q}_\mathrm{e}\cdot\boldsymbol{\nu}
= \frac{1-r_\mathrm{e}}{1+r_\mathrm{e}}
\Bigl(\left| w_\mathrm{e} \tilde{\mathbf{v}}_{\mathrm{d},\mathrm{e}}\cdot\boldsymbol{\nu} \right|
+\frac{2}{3} w_\mathrm{e} v_{\mathrm{th},\mathrm{e}} 
\Bigr) 
-\frac{2}{1+r_\mathrm{e}}
\gamma u_\mathrm{e}^\gamma
\sum_i\max(\mathbf{\Gamma}_i\cdot\boldsymbol{\nu},0)\,,
\label{eq:boundary_eps}
\end{eqnarray}
where $\boldsymbol{\nu}$ represents the normal vector pointing toward the plasma boundaries, and $r_\mathrm{e}$, $\gamma$, $u_\mathrm{e}^\gamma$ and $\mathbf{\Gamma}_i$ denote the electron reflection coefficient, the secondary electron emission coefficient, mean energy of secondary electrons and the ion fluxes at the boundaries, respectively. 
The vector of electron drift velocity $\mathbf{v}_{\mathrm{d},\mathrm{e}}$, the thermal velocity of electron $v_{\mathrm{th},\mathrm{e}}$, and the vector of electron energy drift velocity $\tilde{\mathbf{v}}_{\mathrm{d},\mathrm{e}}$ in the case of the conventional drift-diffusion approximation (option \texttt{DDAc}) are defined as
\begin{eqnarray}
\mathbf{v}_{\mathrm{d},\mathrm{e}} 
= -b_\mathrm{e} \mathbf{E}\, ,
\qquad
%\\
v_{\mathrm{th},\mathrm{e}} 
= \sqrt{\frac{8 k_\mathrm{B} T_\mathrm{e}}{\pi m_\mathrm{e}}}\, ,
\qquad
%\\
 \tilde{\mathbf{v}}_{\mathrm{d},\mathrm{e}} 
= -\tilde{b}_\mathrm{e} \mathbf{E}\, ,
\label{eq:V_drift_DDAc}\
\end{eqnarray}
while in the case of commonly used drift-diffusion approximation (option \texttt{DDA53}) the vector of electron energy drift velocity $\tilde{\mathbf{v}}_{\mathrm{d},\mathrm{e}}$ is equal to $-\frac{5}{3}b_\mathrm{e}\mathbf{E}$. For the improved drift-diffusion approximation (option \texttt{DDAn}), $\mathbf{v}_{\mathrm{d},\mathrm{e}}$ and $\tilde{\mathbf{v}}_{\mathrm{d},\mathrm{e}}$ are defined differently, taking the expressions
\begin{eqnarray}
\mathbf{v}_{\mathrm{d},\mathrm{e}} 
= -\frac{e_0}{m_\mathrm{e}\nu_\mathrm{e}} \mathbf{E}\, ,
\qquad
%\\
 \tilde{\mathbf{v}}_{\mathrm{d},\mathrm{e}} 
= -\frac{e_0}{m_\mathrm{e}\tilde{\nu}_\mathrm{e}}\Bigl(\frac{5}{3}
+ \frac{2}{3}\frac{\xi_2}{\xi_0}\Bigr) \mathbf{E}\, .
\qquad
\label{eq:V_drift_DDAn}\     
\end{eqnarray}
In relations (\ref{eq:V_drift_DDAc}) and (\ref{eq:V_drift_DDAn}), $T_\mathrm{e} = 2u_\mathrm{e}/(3k_\mathrm{B})$ is the temperature of electrons and $k_\mathrm{B}$ is the Boltzmann constant. 

The boundary condition for the balance equation of  heavy particle densities has the following form

\begin{equation}
\mathbf{\Gamma}_\mathrm{h}\cdot\boldsymbol{\nu}
= \frac{1-r_\mathrm{h}}{1+r_\mathrm{h}}
\Bigl(\left|\mathrm{sgn}(q_\mathrm{h}) 
n_\mathrm{h} \mathbf{v}_{\mathrm{d},\mathrm{h}}  \cdot\boldsymbol{\nu}\right|
+\frac{1}{2} n_\mathrm{h} {v}_{\mathrm{th},\mathrm{h}} \Bigr),     
\label{eq:boundary_heavy}\\
\end{equation}
where the variables and coefficients associated to heavy particles are defined in a manner analogous to that of the electrons. 

For the Poisson equation (\ref{eq:poisson}), the boundary conditions are defined by setting the applied voltage $U_\mathrm{a}$ at the powered electrode and zero potential at the grounded electrode.
The accumulation of surface charges is additionally taken into account in the case of dielectric boundaries, as described in~\cite{Stankov_2020}, using the interface condition
\begin{equation}
- \varepsilon_\mathrm{r}
\varepsilon_0
\mathbf{E} \cdot \boldsymbol{\nu} 
=
\sigma.
\label{eq:boundary_charge}
\end{equation}
Here, $\varepsilon_\mathrm{r}$ is 1 in the plasma region, while in the dielectric region it has a value characteristic of the considered dielectric. The surface charge density $\sigma$ is determined from the charged particle currents coming
onto the dielectric via equation
\begin{equation}
\frac{\partial \sigma}{\partial t}
=
\sum_{j}q_j
\mathbf{\Gamma}_j \cdot \boldsymbol{\nu} .
\label{eq:boundary_sigma}
\end{equation}

\subsubsection{Source terms}
\label{source_terms}

The source terms \( S_j \) of equations (\ref{eq:continuity}) is defined as 
\begin{eqnarray}
S_j = \sum_{l=1}^{N_\mathrm{r}} (G_{jl} - L_{jl}) R_l, 
\label{eq:S_j}\,
\end{eqnarray}
where \( R_l \) is the reaction rate of reaction \( l \), given by
\begin{equation}
R_l = k_l \prod_{i=1}^{N_\mathrm{s}} n_i^{\beta_{il}}.
\label{eq:R_l}
\end{equation}
Here, \( \beta_{il} \) and \( k_l \) denote the partial reaction order of species \( i \) and the rate coefficient for reaction \( l \). \( N_\mathrm{r} \) represents the number of reactions, while \( N_\mathrm{s} \) is the number of species considered in the model.
In relation~(\ref{eq:S_j}), \( G_{jl} \) and \( L_{jl} \) are the gain and loss matrix elements, respectively. They are defined by the stoichiometric coefficients for the given species and reactions. MCPlas automatically generates these matrices from the RKM input data, which facilitates effortless switching between models with different levels of complexity.

The source terms \( \tilde{S}_e \) of equation~(\ref{eq:we}) is defined as 
\begin{eqnarray}
\tilde{S}_\mathrm{e} = \sum_{l=1}^{N_\mathrm{r}} \tilde{R}_l,
\label{eq:S_e}\,
\end{eqnarray}
where $\tilde{R}_l$ is the electron energy rate for reaction \( l \), which is usually defined as reaction rate \( R_l \) multiplied by the net electron energy change (gain or loss) \( \Delta \varepsilon_l \). 
In the case where electron energy rate coefficient $\tilde{k}_l$ is defined, e.g.\ for elastic collision, $\tilde{R}_l$ is determined as $ \tilde{k}_l \prod_{i=1}^{N_\mathrm{s}} n_i^{\beta_{il}}$. 
For reactions in which electrons do not participate \( \tilde{R}_l \) is equal to zero.

\subsection{Transport and rate coefficients}
\label{transport_and_rate_coefficients}

As described in section~\ref{particle_fluxes}, three different drift-diffusion approximation for the electron component are supported by MCPlas. 
For \texttt{DDAc} and \texttt{DDA53}, electron transport coefficients are typically obtained by solving the stationary, spatially homogeneous Boltzmann equation for prescribed reduced electric field. 
Users may also provide coefficients derived by other theoretical or experimental methods. 
In contrast, \texttt{DDAn} explicitly requires solving the electron Boltzmann equation and defining transport coefficients as integrals of the isotropic part $f_0$ and the first two contributions $f_1$ and $f_2$ to the anisotropy of the EVDF over the kinetic energy $U$ of the electrons, respectively, according to
\begin{eqnarray}
\nu_\mathrm{e} = \frac{2}{3m_\mathrm{e}\mathit{\Gamma}_\mathrm{e}}\int\limits_0^\infty 
\frac{U^{\nicefrac{3}{2}}}{\lambda_\mathrm{e}(U)} f_1(U)\,\mathrm{d}U\,, \label{eq:nu}\\
\tilde{\nu}_\mathrm{e} = \frac{2}{3m_\mathrm{e} Q_\mathrm{e}}\int\limits_0^\infty 
\frac{U^{\nicefrac{5}{2}}}{\lambda_\mathrm{e}(U)} f_1(U)\,\mathrm{d}U\,, \label{eq:enu}\\
\xi_0 = \frac{2}{3 n_\mathrm{e}}\int\limits_0^\infty 
U^{\nicefrac{3}{2}} f_0(U)\,\mathrm{d}U\,, \label{eq:transp_first}\\
\xi_2 = \frac{4}{15 n_\mathrm{e}}\int\limits_0^\infty 
U^{\nicefrac{3}{2}} f_2(U)\,\mathrm{d}U\,, \\	
\tilde{\xi}_0 = \frac{2}{3 w_\mathrm{e}} \int\limits_0^\infty 
U^{\nicefrac{5}{2}} f_0(U)\,\mathrm{d}U\,, \\
\tilde{\xi}_2 = \frac{4}{15 w_\mathrm{e}} \int\limits_0^\infty 
U^{\nicefrac{5}{2}} f_2(U)\,\mathrm{d}U\,. \label{eq:transp_last}		
\end{eqnarray}

It should be noted that MCPlas does not provide built-in computation of electron transport and rate coefficients. In particular, no Boltzmann solver is included, and the toolbox does not implement direct coupling between the electron Boltzmann equation and the fluid-Poisson model. Therefore, the required coefficients must be determined externally using any preferred approach and provided as an input by the user. The same applies to coefficients associated with heavy particle species. Users need to ensure that the provided input data are defined in accordance with the description given in section~\ref{reaction_kinetics_model}.  
\subsection{Stabilisation techniques}
To ensure numerical stability and obtain consistent solutions, MCPlas provides the possibility to use some stabilisation techniques.
As an option, the toolbox offers applying the logarithmic transformation of the densities of particle species and the mean electron energy. 
This transforms the general balance equation 
\begin{equation}
 \frac{\partial P}{\partial t}
        + \nabla\cdot\mathbf{F} = S\,,
        \label{eq:general_balance}\\
\end{equation}
via the relations
\begin{eqnarray}
 p=\mathrm{ln}(P)\,,
 \label{eq:log_form}\, \\
 \frac{\partial P}{\partial t}
 = P \frac{1}{P} \frac{\partial P}{\partial t}
 = P \frac{\partial \mathrm{ln}(P)}{\partial P} \frac{\partial P}{\partial t}
 = P \frac{\partial \mathrm{ln}(P)}{\partial t}
 = P \frac{\partial p}{\partial t}\,,
 \label{eq:transform}   
\end{eqnarray}
into the form
\begin{equation}
 \mathrm{e}^p\frac{\partial p}{\partial t}
        + \nabla\cdot\mathbf{F} = S\,.
        \label{eq:general_balance_transformed}\\
\end{equation}
Here, $P$ stands for $n_j$ and $w_\mathrm{e}$, $\mathbf{F}$ for $\mathbf{\Gamma}_j$ and $\mathbf{Q}_\mathrm{e}$, and $S$ for $S_j$ and  $\tilde{S}_\mathrm{e}$. 
This approach inherently enforces positivity of the solution and suppresses oscillations in regions with steep gradients or low concentrations.

The toolbox also implements a source term stabilisation method applicable to all particles and electron energy balance equations, in a similar way to that used in the CPM~\cite{CPM_user_guide}. 
When enabled, an additional term  
\begin{equation}
N_A \exp(-\xi \ln(p))
\end{equation}
is added to the right-hand side of the balance equations, where $p$ is the representative solution quantity. Here, $N_\mathrm{A}$ denotes the Avogadro constant, $\xi\in[0.25,1]$ is a stabilisation parameter (default value is 1). 
This term is designed to counteract numerical instabilities caused by stiff or nonlinear reactions. 
It acts as a buffer at very low particle number densities, preventing the appearance of negative values, and becomes negligible at higher number densities.
Together, these techniques significantly improve the robustness and physical consistency of plasma simulations, enabling reliable modelling across diverse conditions.

\subsection{Code structure and user interface}

\begin{figure}
	\centering
	\includegraphics[]{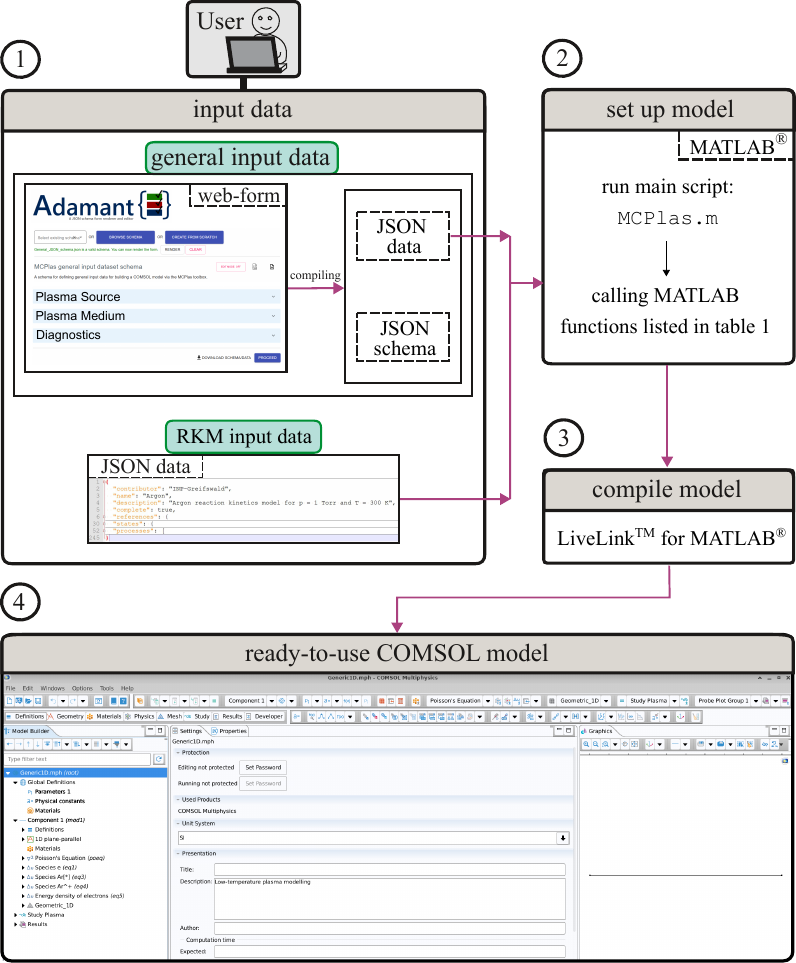}
	\caption{Illustration of the MCPlas workflows including the provision of input data (1), the processing (2), (3) and the implemented plasma model in COMSOL (4).}
	\label{fig:figure_2}
\end{figure}

 The general concept of the MCPlas toolbox is given by the workflow presented in figure~\ref{fig:figure_2}.
 At the beginning of the MCPlas workflow, all input data necessary for setting up the model must be provided. 
 This considers general and RKM input data prepared in JSON data format as described in subsection~\ref{standardised_input_schemas}.
The second step of the MCPlas workflow involves setting up the model by executing the main MATLAB script, \texttt{MCPlas.m}. 
This script systematically calls MATLAB functions specifically designed for the toolbox. The list of these functions with their description is given in table~\ref{table_matlab_functions}.
\begin{table}[htbp]
\centering
\caption{List of MATLAB functions designed for the MCPlas toolbox in order of their execution.}
\begin{tabular}{lp{7cm}}  % 'l' for function name, 'p{10cm}' for wrapped description
\toprule
MATLAB functions & Description \\
\midrule
\texttt{MCPlas.m} & main MATLAB file of the toolbox \\
\texttt{ReadJSON.m} & read JSON input files \\
\texttt{InpRKM.m} & define variables from RKM input data \\
\texttt{InpGeneral.m} & define variables from general input data \\
\texttt{SetParameters.m} & set global parameters for the model\\
\texttt{SetGeometry.m} & set modelling geometry (see section~\ref{modelling_geometries}) \\
\texttt{SetConstants.m} & set constants \\
\texttt{SetVariable.m} & set model variables \\
\texttt{SetTransportCoefficients.m} & set transport coefficients (see section~\ref{transport_and_rate_coefficients}) \\
\texttt{SetRateCoefficients.m} & set rate coefficients (see section~\ref{transport_and_rate_coefficients}) \\
\texttt{SetEnergyRateCoefficients.m} & set electron energy rate coefficients  \\
\texttt{SetRates.m} & set rates for all included reactions \\
\texttt{SetEnergyRates.m} & set electron energy rates for all electron-involving reactions \\
\texttt{SetFluxes.m} & set fluxes (see section~\ref{particle_fluxes}) \\
\texttt{SetSources.m} & set source terms for fluid equations (see section~\ref{source_terms}) \\
\texttt{AddSurfaceChargeAccumulation.m} & add surface charge accumulation equation if required (see section~\ref{boundary_fluxes}) \\
\texttt{AddPoissonEquation.m} & add Poisson equation  (see section~\ref{fluid_poisson_model}) \\
\texttt{AddFluidEquations.m} & add fluid equations for included species and mean electron energy (see section~\ref{fluid_poisson_model}) \\
\texttt{SetElectrical.m} & set variables for electrical properties \\
\texttt{SetProbesAndGraphs.m} & set probes and graphs for post-processing \\
\texttt{SetMesh.m} & set computational mesh \\
\texttt{SetProject.m} & set plasma modelling study, solver sequences and numerical parameters \\
\bottomrule
\end{tabular}
\label{table_matlab_functions}
\end{table}
The third step considers employing LiveLink\textsuperscript{\texttrademark} \textit{for} MATLAB\textsuperscript{\fontsize{4}{4}\selectfont\textregistered} to compile the COMSOL model. The second and third step automatically execute after running the MCPlas code, i.e. they are not the user's concern. Finally, in the fourth step, the COMSOL model is saved and ready to use. Once the model has been built, it behaves like any other COMSOL model. It retains all standard COMSOL operations, allowing the user to, for example, adjust model details that are problem-specific, perform parametric studies by changing parameters directly in COMSOL, modify or extend the model structure for more advanced scenarios, etc.

\subsection{Standardised input schemas}
\label{standardised_input_schemas}

All input data required by MCPlas are provided in structured JSON format. This concerns the general input specifying the setup as well as the RKM. To organise and validate these inputs, MCPlas relies on standardized JSON schemas that define all required fields. Using these schemas reduces the risk of errors and ambiguities while simplifying automated parsing and model generation.

\subsubsection{Reaction kinetics model}
\label{reaction_kinetics_model}

\begin{figure}
	\centering
	\includegraphics[width=\textwidth]{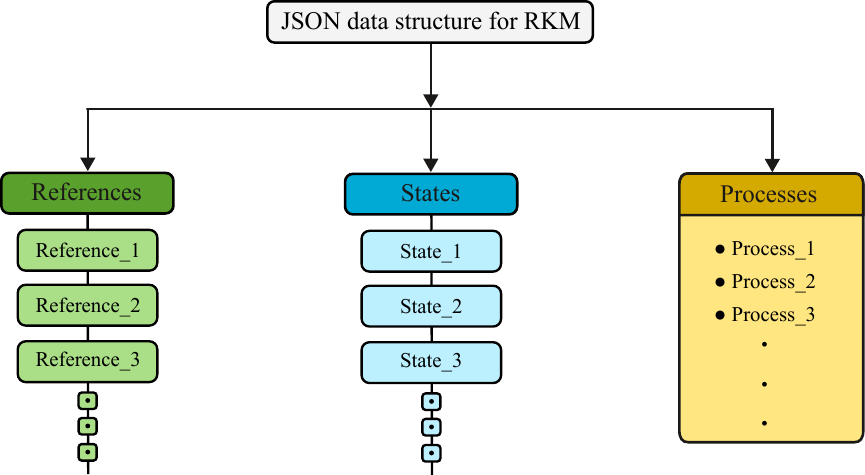}
	  \caption{Schematic representation of the top-level structure of an LXCat JSON document for LTP input data.}
	\label{fig:figure_3}
\end{figure}

Ideally, MCPlas would use a standardised format to define its RKM. However, a comprehensive standard for plasma chemistry data does not yet exist. A practical alternative is to adopt and build on current work aimed at achieving these goals. To this end, MCPlas adopts an extended version of the format proposed for the next generation of the LXCat platform, which is currently under active open-source development~\cite{LXcat_paper_2021, lxcat_github}. The top-level structure of the JSON data describing an RKM according to the schema introduced by LXCat is shown in figure~\ref{fig:figure_3}. It comprises three main properties: \texttt{references}, an object storing the references from which included data are extracted; \texttt{states}, an object listing properties of the all states/particle species included in the model; and \texttt{processes}, an array of process objects providing information on the reaction equations and corresponding data. Each individual element of the JSON document is formaly defined by its corresponding JSON schema definition. The exact definition of the LXCat schemas can be found in the LXCat GitHub repository~\cite{lxcat_github}. These schemas can also be used to validate incoming documents. The development of MCPlas has contributed to the extension of existing electron scattering schemas to further accommodate plasma chemistry data.

\begin{figure}
	\centering
	\includegraphics[width=\textwidth]{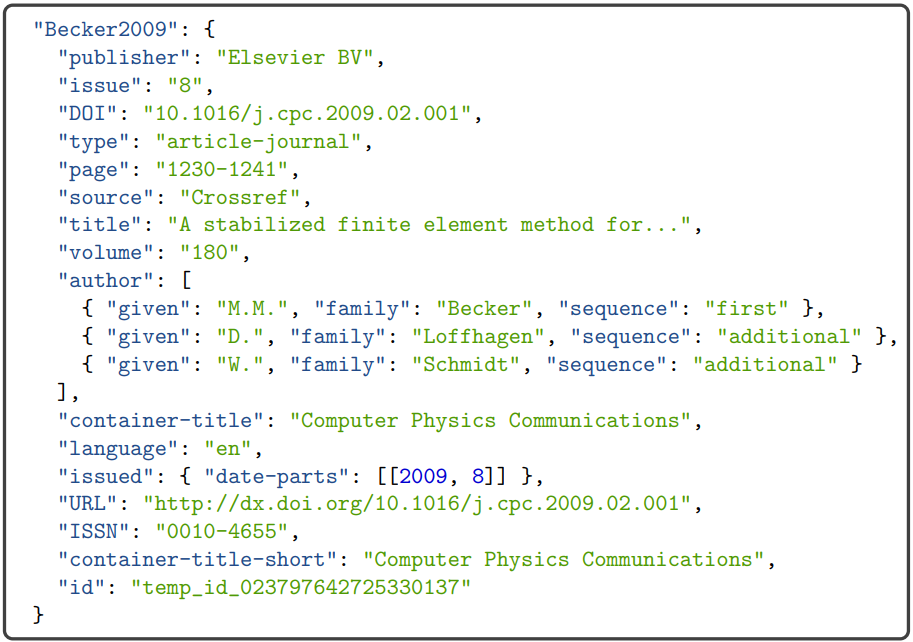}
	  \caption{Example of a JSON object representing a reference, as defined by the CSL-JSON schema.}
	\label{fig:figure_4}
\end{figure}

Complex RKMs incorporate data from different sources provided by numerous authors. It can be challenging to keep track of all required references to used data. However, proper referencing is crucial in an academic setting for the sake of facilitating traceable and reproducible results. 
Therefore, the top-level property \texttt{references} stores detailed definitions of all used references using the standardised
Citation Style Language JSON (CSL-JSON) format~\cite{csl_json}. The main advantage of adopting the CSL-JSON format is its established ecosystem, which includes a schema definition and tools that allow, for example, retrieval of CSL-JSON references by DOI, and automatic conversion to and from different citation formats such as BibTeX and RIS. To avoid repetition of complex reference definitions, the keys of the \texttt{references} object act as unique identifiers that are used elsewhere in the document. An example of a JSON object representing a reference is displayed in figure~\ref{fig:figure_4}. 
More details on the CSL-JSON format can be found in~\cite{Master_Thesis_Boer}.

\begin{figure}
	\centering
	\includegraphics[width=\textwidth]{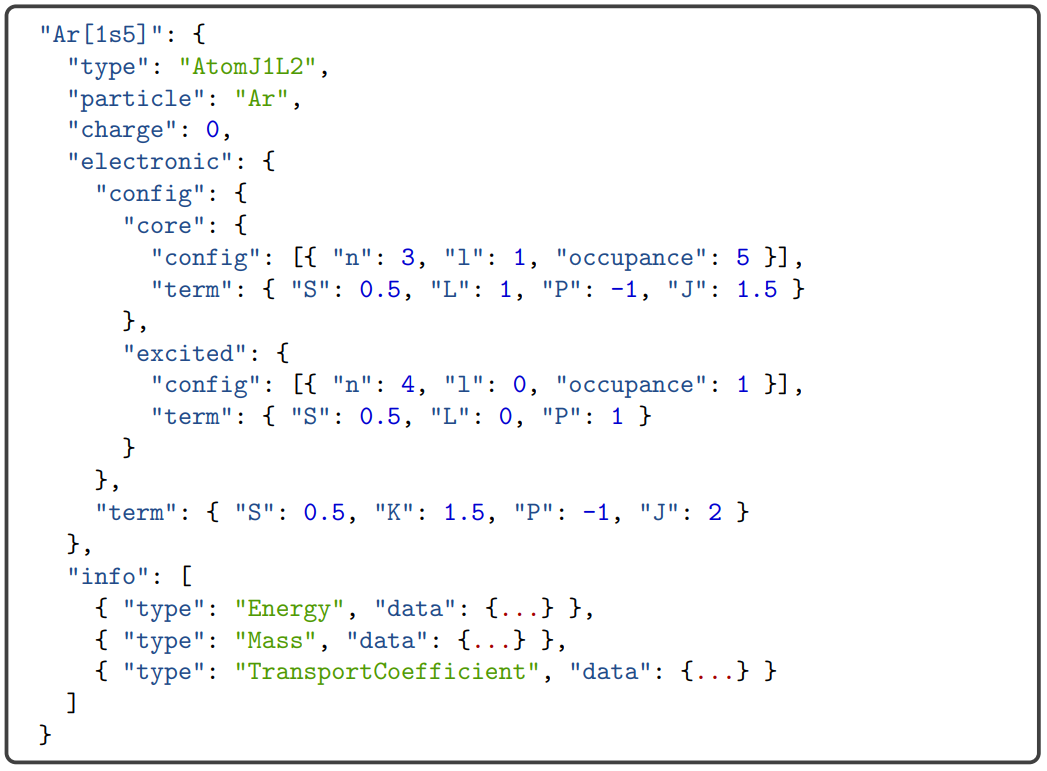}
	  \caption{Example of a key-value pair from the \texttt{states} property. The JSON object defines the argon $\mathrm{1s5}$ metastable state using the $\mathrm{J_1L_2}$ coupling scheme. The \texttt{info} property stores an array of different data related to the species, some possible options are \texttt{Mass}, \texttt{Energy}, \texttt{Mobility}, and \texttt{DiffusionCoefficient}. The \texttt{Ar[1s5]} key can be used to reference this state in other parts of the document.}
	\label{fig:figure_5}
\end{figure}

Similar to the \texttt{references} property, the value of the \texttt{states} top-level property is an object whose values include detailed descriptions of the involved species following a well-defined schema. As an example, the JSON object representing the $\mathrm{1s_5}$ excited state of the argon atom is given in figure~\ref{fig:figure_5}. 
A \texttt{states} object stores the type of the \texttt{particle}, its \texttt{charge}, and a description of the excited level that the species resides in. This example uses the $\mathrm{J_1L_2}$ coupling scheme to describe the electronic configuration of the excited state. Many different types of state descriptions are supported, both for atoms and various types of molecules. Moreover, the schema can be easily extended to support additional species types. The philosophy behind the species schema design, and its practical implementation are highlighted in~\cite{Master_Thesis_Boer}. The latest implementation and schema definitions can be found in the LXCat GitHub repository~\cite{lxcat_github}.

\begin{figure}
	\centering
	\includegraphics[width=\textwidth]{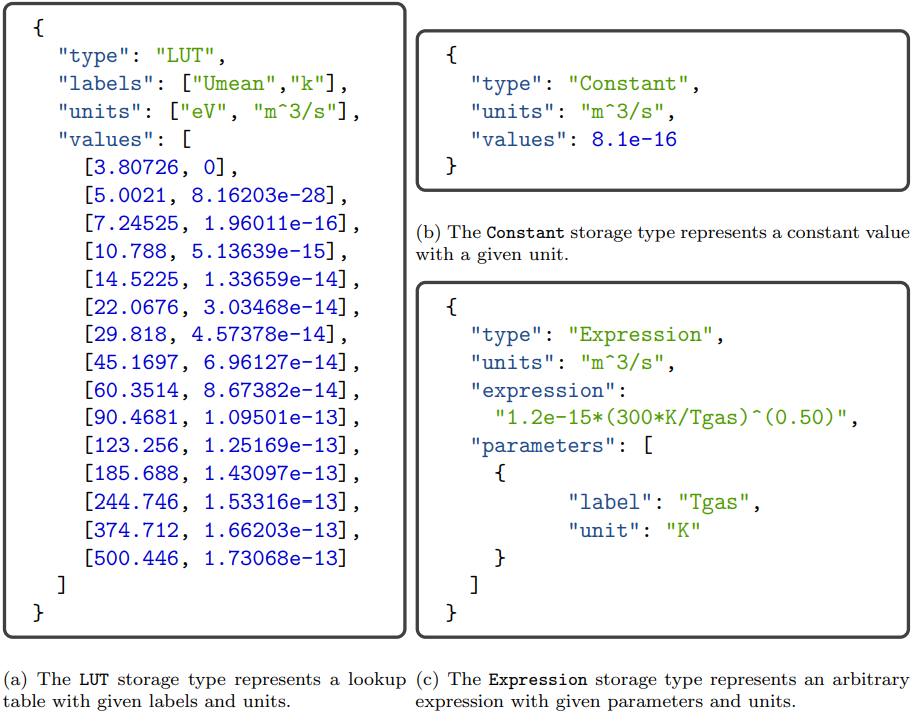}
	  \caption{Examples of JSON objects for the different supported data storage types.}
	\label{fig:figure_6}
\end{figure}

In addition to the state description, a state object also hosts characteristic data related to the species in its \texttt{info} property. This property stores an array of data objects of different types. Some examples of possible types are \texttt{Mass}, \texttt{Energy} and \texttt{TransportCoefficient}. An entry in the \texttt{info} array can store its accompanying data using multiple different storage types. The currently supported storage types include: \texttt{LUT}, a lookup table; \texttt{Constant}, a constant value; and \texttt{Expression}, an arbitrary algebraic expression that depends on the specified plasma parameters. Figure~\ref{fig:figure_6} shows three examples of \texttt{data} objects, one for each of the currently supported storage types.

\begin{figure}
	\centering
	\includegraphics[width=\textwidth]{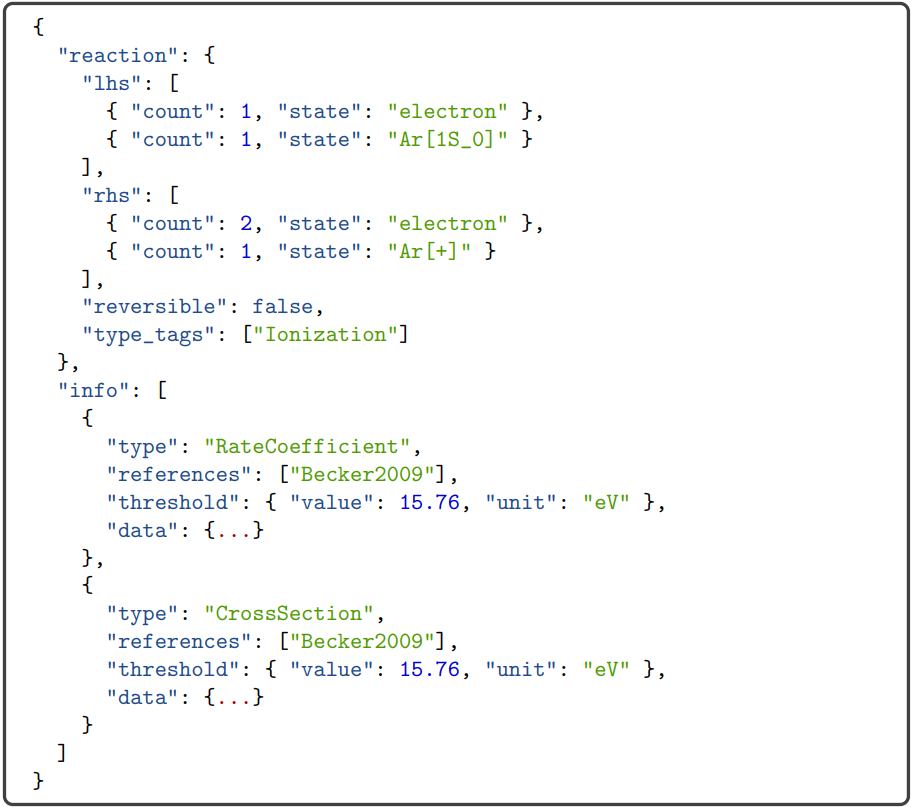}
	  \caption{Example of a JSON object entry in the \texttt{processes} property. This object defines an electron-impact ionisation reaction from the argon ground state: $\mathrm{e}^- + \mathrm{Ar\left(^1S_0\right)} \rightarrow 2\mathrm{e}^- + \mathrm{Ar}^+$. The \texttt{reaction} property stores the basic definition of the reaction. The \texttt{electron}, \texttt{Ar[1S\_0]} and \texttt{Ar[+]} strings are keys in the \texttt{states} object that reference the electron, argon ground state, and the argon ion, respectively. The \texttt{info} property stores an array of different data related to the process. Similar to the state keys, the \texttt{Becker2009} string is a key in the \texttt{references} object.}
	\label{fig:figure_7}
\end{figure}

The final top-level property of interest is the \texttt{processes} property. The value of this property is an array of process objects that define the collisional and radiative processes considered. An example of an object describing electron-impact ionisation of the ground state argon atom is given in figure~\ref{fig:figure_7}. Each  \texttt{process} object consists of two properties, \texttt{reaction} and \texttt{info}. The former defines general information about the reaction. Most notably, the \texttt{lhs} (left-hand side) and \texttt{rhs} (right-hand side) properties specify the reactants and products, respectively. Note that state definitions are referenced by their corresponding keys in the top-level \texttt{states} object. Data objects related to the process are stored in its \texttt{info} array. This property behaves similarly to its state object counterpart. Examples of supported data types for processes include \texttt{CrossSection}, \texttt{RateCoefficient}, and \texttt{EnergyRateCoefficient}. The latter is meant to be used when the electron energy rate cannot be determined by simply multiplying the corresponding threshold energy and reaction rate, e.g. for elastic collisions or electron-ion recombination reactions. References can be linked by adding their corresponding key in the top-level \texttt{references} object to the \texttt{references} array of the info object. Finally, the data can be represented using one of the available data storage types highlighted in figure~\ref{fig:figure_6}.

\subsubsection{General input data}

In addition to the input data specifying the RKM and the species transport properties, general input data defining the setup are required to build the model. 
These general input data are provided in the JSON format as well and include information on plasma source, plasma medium, and diagnostics method. 
The plasma source field describes the geometry, electrical, and material properties of the source. 
The plasma medium field encompasses the general characteristics of the gas under study, as well as the surface properties specific to the included species and surface materials. 
Finally, the diagnostics field contains the relevant properties of the fluid-Poisson model, which is employed here as a diagnostic tool for investigation.

For the preparation of the general input data in JSON data format, employing the Adamant tool~\cite{Adamant_2022} for collection of JSON schema-based metadata is proposed. 
This tool is primarily intended to facilitate the implementation of digital research data management processes by enabling easy compilation and creation of metadata and metadata schemas based on JSON schema standards. All the features of the Adamant are very convenient for generating the JSON data format containing all general input data necessary for model building.
In general, Adamant can generate JSON data files based on the included JSON schema. 
The JSON schema can be included in three ways: (i) selecting one of the existing schemas, (ii) uploading a schema already prepared by the user, or (iii) creating a schema from scratch directly on the platform. 
The MCPlas toolbox comes with a prepared JSON schema to collect the required general input data based on Plasma-MDS~\cite{PLASMA_MDS}, a metadata schema for plasma science.
The user has to upload the provided JSON schema to the Adamant platform and start the rendering process. 
Subsequently, Adamant automatically generates an interactive web-form, whose elements correspond to the general input data that the user has to complete.
Compiling the fully defined web-form generates a JSON data file containing all general input data needed for the model building with MCPlas. 
If the user wants to make a modelling analysis with changed general input data, they just need to generate a modified JSON data file.
For the purposes of MCPlas, Plasma-MDS was specifically extended to correspond to the general input data required to set up the fluid-Poisson model in COMSOL. 
With this, the procedure is designed to promote the further implementation of the FAIR data principles to plasma modelling.

\section{Results}
\label{sec:results}

The results provided in this section intend to emphasise three significant aspects concerning the advantages of employing the MCPlas toolbox. The first one pertains to the importance of the improved description of electron transport and boundary conditions described in section~\ref{basic_equations}. In this regard, the model generated by MCPlas is compared with CPM, one of the most widely used commercial software in the plasma modelling community. The second aspect relates to the ability of MCPlas to manage complex RKMs. This is illustrated by the example of the 23-species argon RKM introduced in~\cite{Stankov_2020}. The third aspect concerns the reusability of JSON input data defining the RKM in different plasma modelling toolkits. For that purpose, test cases implemented with MCPlas were computed by two additional plasma modelling toolboxes: PLASIMO~\cite{plasimo, vanDijk-2009-ID2562} and FEDM (Finite Element Discharge Modelling code)~\cite{Jovanovic-2023-ID6156,Jovanovic_Finite_Element_Discharge}.

\begin{figure}
	\centering
	\includegraphics[]{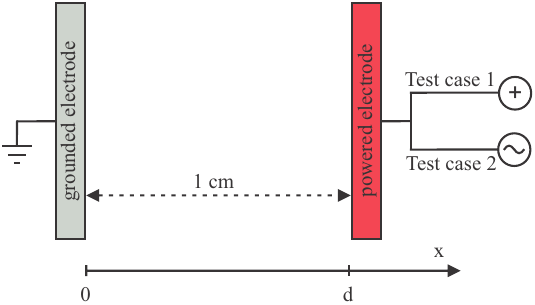}
	\caption{Sketch of the discharge geometry used in the modelling studies with DC (test case 1) and RF (test case 2) applied voltage.}
	\label{fig:figure_8}
\end{figure}

All modelling studies were performed for two test cases, namely DC and RF low-pressure glow discharges in argon. Figure~\ref{fig:figure_8} shows a schematic representation of the geometry used in the spatially one-dimensional modelling studies (\texttt{1D} option). In both test cases, the left electrode is grounded whereas the right one is powered. In the case of DC discharge, a constant voltage of $\unit[350]{V}$ is applied, and for the RF case, a sinusoidal voltage with an amplitude of $\unit[350]{V}$ and a frequency of $\unit[13.56]{MHz}$ is used. The gap distance, gas pressure and gas temperature are set to $\unit[1]{cm}$, $\unit[1]{Torr}$ and $\unit[300]{K}$, respectively. %For all modelling cases modelling domain is subdivided into 1000 equidistant mesh elements. A relative tolerance of $10^{-4}$ is used to control the time step of the solver.

\subsection{Comparison of MCPlas and CPM}
\label{sec:plasmamod}
The improved drift-diffusion approximation for the particle and energy flux of electrons (\texttt{DDAn} in MCPlas) is not applicable in CPM. Furthermore, CPM employs its own boundary condition formulation for the balance equations~\cite[Chapter~4]{CPM_user_guide}. Thus, a direct comparison of MCPlas and CPM modelling outcomes can reveal the significance of MCPlas features related to the description of the electron transport and boundary conditions.  In that regard, a time-dependent, spatially one-dimensional modelling approach utilising a 4-species argon RKM in both CPM and MCPlas was performed.
The considered RKM takes into account electrons, the argon atom in its ground state $\mathrm{Ar[1p_0]}$, the argon ion $\mathrm{Ar^+}$, and argon atoms in a lumped excited state $\mathrm{Ar^*}$. The lumped excited state represents the two metastable states of the argon atom $\mathrm{Ar[1s_5]}$ and $\mathrm{Ar[1s_3]}$. Electron transport and rate coefficients are pre-calculated from the solution of the spatially homogeneous, steady-state Boltzmann equation for a set of given reduced electric field strengths using the multi-term method described in \cite{LEYH199833}. Note that CPM includes the possibility of using its own Boltzmann solver, but since it is based on the two-term approximation, it was not used in this work to ensure a comparison of the models based on identical input data. More details about about the transport coefficients of the included species and the rate coefficients of the considered chemical reactions can be found in~\cite{Becker_2009}. The values of surface parameters in boundary conditions~\eqref{eq:boundary_e}, \eqref{eq:boundary_eps} and \eqref{eq:boundary_heavy} are set in accordance with the studies given in~\cite{Becker_2009}. 

\subsubsection{Test case 1: DC glow discharge}
\label{subsec:testcase_1}
\begin{figure}
	\centering
	\includegraphics[]{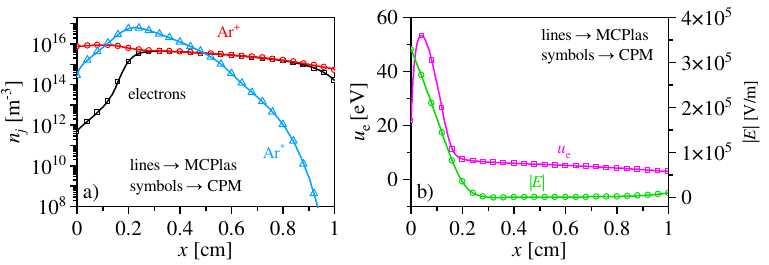}
	\caption{Verification of the MCPlas toolbox for a DC glow discharge in argon. Comparison of species number densities a), and mean electron energy and electric field b) in the stationary state of the discharge obtained with the model built in CPM and the same model generated by the MCPlas toolbox. 
	}
	\label{fig:figure_9}
\end{figure}

For verification of the proposed toolbox implementation, MCPlas was configured to generate the the same plasma model as implemented by the CPM. In particular, the electron and electron energy fluxes were defined according to the relations given in equations (\ref{eq:JeDDAc}) and (\ref{eq:QeDDA53}) (\texttt{DDA53}), while the boundary conditions for the basic model equations were adapted to be the same as the default ones in CPM.
Figure~\ref{fig:figure_9} presents modelling results for the stationary state of the DC discharge in argon obtained by the equation-based plasma model implemented using MCPlas and the plasma model provided by CPM. The particles number densities as well as the mean electron energy and electric field calculated by both models are in perfect agreement. Furthermore, both models successfully reproduce the behaviour of a stable glow discharge at low pressure described in the literature~\cite{RaizerBook}. In particular, the cathode-sheath region, characterised by a high, linearly increasing electric field near the cathode ($x=\unit[0]{}$), is clearly evident from the modelling results. This excellent agreement between the both models indicates that the proposed toolbox is capable of generating correct models.

\begin{figure}
	\centering
	\includegraphics[]{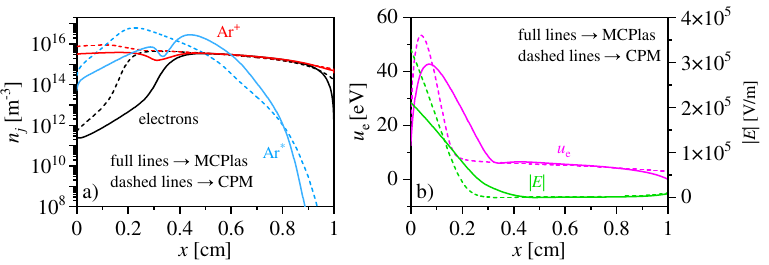}
	\caption{Comparison of particle number densities a) as well as mean electron energy and electric field b) in the stationary state of a DC glow discharge in argon obtained with the model generated by MCPlas toolbox (option \texttt{DDAn} and boundary conditions (\ref{eq:boundary_e}), (\ref{eq:boundary_eps}) and (\ref{eq:boundary_heavy})) and the model built in CPM.
	}
	\label{fig:figure_10}
\end{figure}

After this verification step, the MCPlas toolbox was tested for the same discharge conditions with inclusion of the novel drift-diffusion approximation (option \texttt{DDAn}) and boundary conditions (\ref{eq:boundary_e}), (\ref{eq:boundary_eps}) and (\ref{eq:boundary_heavy}). 
The comparison of the modelling results obtained from the model generated by the MCPlas toolbox and the model built in CPM is displayed in figure~\ref{fig:figure_10}. 
Figure~\ref{fig:figure_10} a) shows that the calculated number densities of all species show distinct differences across the entire gap. This refers especially to the cathode-sheath region where number densities obtained by the MCPlas model are notably lower than those calculated by the CPM model. 
The results for $u_\mathrm{e}$ and $E$ presented in figure~\ref{fig:figure_10} b), apart from magnitude differences, show that the MCPlas model predicts a broader cathode-sheath region. 
These findings highlight that the electron transport and boundary conditions made available with the MCPlas toolbox significantly influence the modelling results. 
Although the  \texttt{DDAn} approach is well-proven in the literature for its accuracy \cite{Stankov_2022,Loffhagen_2020,Becker-2017-ID4159,Baeva-2019-ID5345}, it was not easily accessible. 
MCPlas facilitates the use of individual plasma model implementations in COMSOL, which can improve the reliability of modelling studies in various scenarios.

\subsubsection{Test case 2: RF glow discharge}
\label{subsec:testcase_2}

\begin{figure}
	\centering
	\includegraphics[]{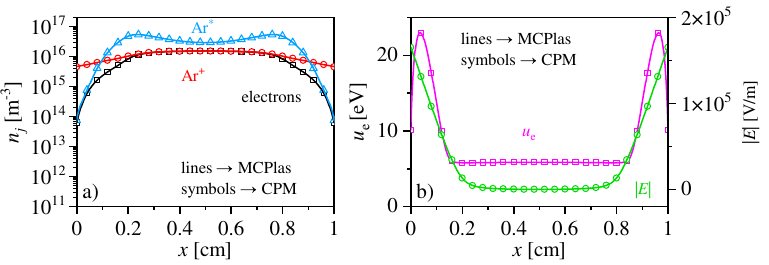}
	\caption{Verification of MCPlas toolbox for an RF glow discharge in argon. Comparison of period-averaged particle number densities a) as well as mean electron energy and electric field b) obtained with the model built in CPM and the same model generated by the MCPlas toolbox for one period in the stable periodic state of discharge. 
	}
	\label{fig:figure_11}
\end{figure}

At first, further verification of the MCPlas toolbox is conducted for an argon RF glow discharge with the same model properties as for the DC glow discharge in section \ref{subsec:testcase_1}. The time-averaged particle number densities, mean electron energy and electric field obtained by the MCPlas and CPM implementations of the same plasma model for one period in the stable periodic state of discharge are presented in figure~\ref{fig:figure_11}. As for the DC discharge case, the proposed toolbox is verified by the excellent agreement of the results of both implementations.

\begin{figure}
	\centering
	\includegraphics[]{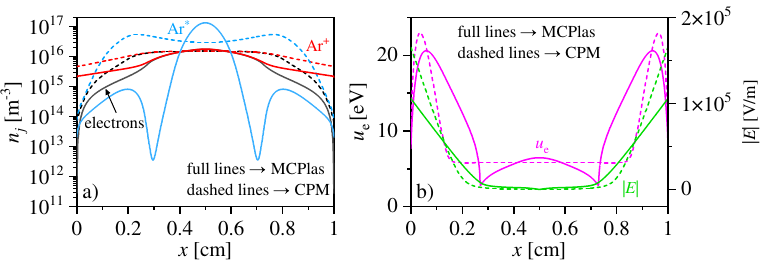}
	\caption{Comparison of period-averaged particle number densities a) and mean electron energy and electric field b) obtained with the model generated by MCPlas toolbox (option DDAn and boundary conditions (10), (11) and (14)) and model built in CPM for one period in the stable periodic state of an RF glow discharge in argon.}
	\label{fig:figure_12}
\end{figure}

The significance of the electron transport description and boundary conditions used for the testing in the DC case was also examined for the RF case. 
Figure~\ref{fig:figure_12} shows a comparison of the time-averaged modelling results for one voltage period of the stable periodic state of the discharge obtained from the model generated by the MCPlas toolbox and the CPM model. 
The differences in the charged species number densities between the models are evident.
As in the DC case, the CPM model predicts higher values than MCPlas model in the cathode-sheath regions.
The spatial profiles of the number density of the excited argon state show significant discrepancies, with computed values differing by over an order of magnitude in the most of the disharge gap (figure~\ref{fig:figure_12} a)). 
The disagreement between the mean electron energy and the electric field calculated by both models, is also clearly notable (figure~\ref{fig:figure_12} b)).

It can be conclude from the outcomes of the two test cases that the difference between MCPlas and CPM models is evident.
Considering that the largest discrepancies are found in the regions near the electrodes, the features of the MCPlas toolbox that are not part of CPM may be of significant relevance also for discharge types where the plasma properties at the boundaries play a key role in the overall plasma dynamics, such as in dielectric barrier discharges.         
   
\subsection{Comparison of different RKMs}
\label{sec:RKMs_comparison}

Building plasma models with a few particle species and simple reaction kinetics is usually a manageable process, regardless of the software used. 
Serious challenges arise with more complex RKMs involving a large number of species and an extended set of collision processes~\cite{Jovanovic_2021,Jovanovic-2023-ID6156}. 
With schema-based input data, the MCPlas toolbox enables the automated generation of models including RKMs of any complexity.
In order to showcase this, modelling of the DC and RF argon glow discharges with the model built via MCPlas using the \texttt{DDAn} approach and boundary conditions (\ref{eq:boundary_e}), (\ref{eq:boundary_eps}) and (\ref{eq:boundary_heavy}) was performed for the same discharge conditions as in subsections~\ref{subsec:testcase_1} and~\ref{subsec:testcase_2}, but with an extended RKM for argon plasmas~\cite{Stankov_2022}. 
\begin{figure}
	\centering
	\includegraphics[]{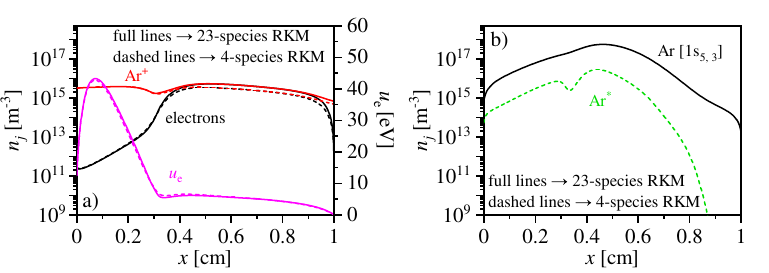}
	\caption{Comparison of number densities of electrons and atomic argon ions, as well as mean electron energy a), and number densities of metastable argon states b) obtained with the model generated by the MCPlas toolbox with 23-species and 4-species RKMs in the stationary state of the DC glow discharge.}
	\label{fig:figure_13}
\end{figure}
\begin{figure}
	\centering
	\includegraphics[]{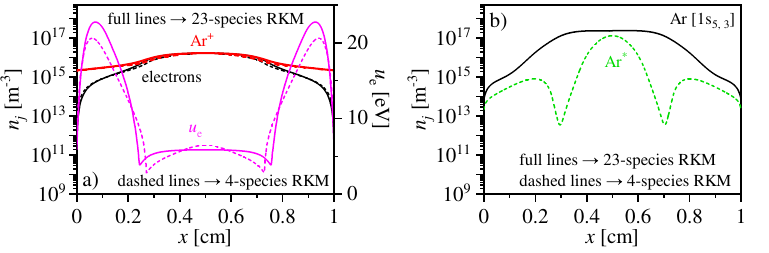}
	\caption{Comparison of period-averaged number densities of electrons and $\mathrm{Ar^+}$ as well as mean electron energy a), and number densities of metastable argon states b) obtained with model generated by MCPlas toolbox with 23-species and 4-species RKMs for the stable periodic state of the RF discharge.
	}
	\label{fig:figure_14}
\end{figure}
The applied argon model considers 23 particle species, including electrons, the ground state atom, the atomic and molecular ion, four excited molecular states ($\mathrm{Ar^*_2[^3\Sigma_u^+,v=0],Ar^*_2[^1\Sigma_u^+,v=0], Ar^*_2[^3\Sigma_u^+,v\gg0], Ar^*_2[^1\Sigma_u^+,v\gg0]}$) and all individual 1s ($\mathrm{Ar[1s_{5,. .,2}]}$) and 2p ($\mathrm{Ar[2p_{10,...,1}]}$) states. 
In addition, energetically higher states are considered in the model as a single lumped state ($\mathrm{Ar^*[hl]}$). 
All species are involved in 409 collision processes and radiative transitions. Similar to the 4-species RKM, electron rate and transport coefficients are determined by solving the Boltzmann equation using the procedure described in \cite{LEYH199833}. All details of this 23-species RKM are given in \cite{Stankov_2022}. 
In order to generate a model using the MCPlas toolbox, all data concerning the species properties and collision processes considered, together with the corresponding references, are prepared in the JSON data format as described in section~\ref{standardised_input_schemas}. The results are then compared with those obtained using the 4-species RKM in figures~\ref{fig:figure_13} and~\ref{fig:figure_14}.

Figure~\ref{fig:figure_13} a) shows that in the DC discharge case, the electron and $\mathrm{Ar^+}$ number densities predicted by the simplified RKM exhibit the same spatial profiles as those of the extended model, differing only slightly in magnitude outside the cathode-sheath regions. The discrepancies are significant for the excited states densities shown in Figure~\ref{fig:figure_13} b). The density of $\mathrm{Ar^*}$ obtained using the 4-species RKM differs by more than an order of magnitude from the summed $\mathrm{Ar[1s_{5}]}$ and $\mathrm{Ar[1s_{3}]}$ levels obtained with the 23-species RKM. For the RF case, the 4-species RKM provides results closely matching those of the extended RKM for charged species (Figure~\ref{fig:figure_14} a)). However, notable deviations appear in the mean electron energy (Figure~\ref{fig:figure_14} a)) and the excited state densities (Figure~\ref{fig:figure_14} b)). 

\begin{figure}
	\centering
	\includegraphics[]{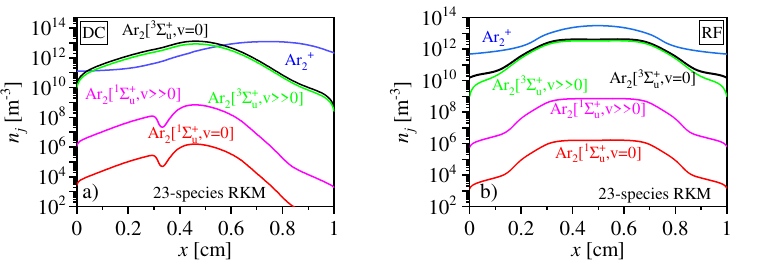}
	\caption{Number densities of excited molecular states and the  molecular ion calculated with the model generated by the MCPlas toolbox with the 23-species RKM in the stationary state of the DC a) and RF b) glow discharges.
	}
	\label{fig:figure_15}
\end{figure}

Clearly, the 23-species RKM, which includes the 15 individual excited states of the argon atom, provides much more information about the plasma composition than the 4-species model, which has only one lumped excited state, $\mathrm{Ar^*}$. Furthermore, the extended model considers additional molecular species that are lacking in the 4-species formulation (Figure~\ref{fig:figure_15}) and thus contributes even more to the understanding of plasma behaviour.

\subsection{Reusability of RKM input data files}
\label{sec:input_reusability}

An important aspect of MCPlas is the reusability of the proposed JSON data format for the complete definition of the RKM input data in different plasma modelling tools. 
Therefore, the reusability of the 4-species and 23-species argon RKM input data used in section~\ref{sec:RKMs_comparison} was tested with two additional sofware packages for LTP modelling. The first one is PLASIMO, a toolbox developed by Plasma Matters~\cite{plasimo} in conjuction with the Eindhoven University of Technology for the numerical simulation of various plasma sources~\cite{vanDijk-2009-ID2562}. Over the years, PLASIMO has gained a strong reputation in the plasma physics community due to its successful application in numerous studies \cite{Hartgers_2001,Hayashi_2004,Mihailova_2008,Maitre_2022}. The second toolbox is FEDM, which was recently developed at the Leibniz Institute for Plasma Science and Technology in Greifswald~\cite{Jovanovic-2023-ID6156}. FEDM is based on the open-source computing platform FEniCS~\cite{Fenics_website} for solving partial differential equations. Both software packages were used to model the DC and RF test cases described in section~\ref{sec:plasmamod} utilising the same JSON schema-based RKM input data. The model setups concerning the treatment of electron transport and boundary conditions were configured to match those used by the CPM and considered for the verification of the MCPlas-generated model in section~\ref{sec:plasmamod}.

\begin{figure}
	\centering
	\includegraphics[]{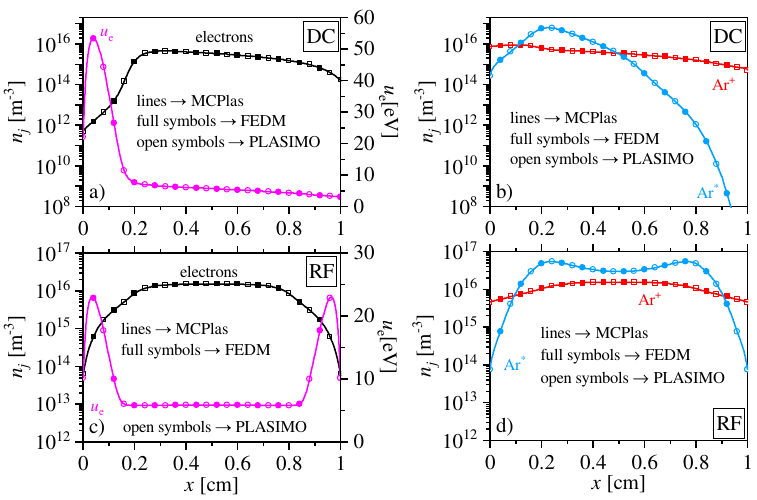}
	\caption{Comparison of modelling results obtained with MCPlas, PLASIMO and FEDM for the 4‑species RKM. The results show the stationary state of the DC (a, b) and the period-averaged state of the RF (c, d) glow discharges in argon for the density and mean energy of electrons (a, c) as well as the heavy-particle densities (b, d).  
 }
	\label{fig:figure_16}
\end{figure}

\begin{figure}
	\centering
	\includegraphics[]{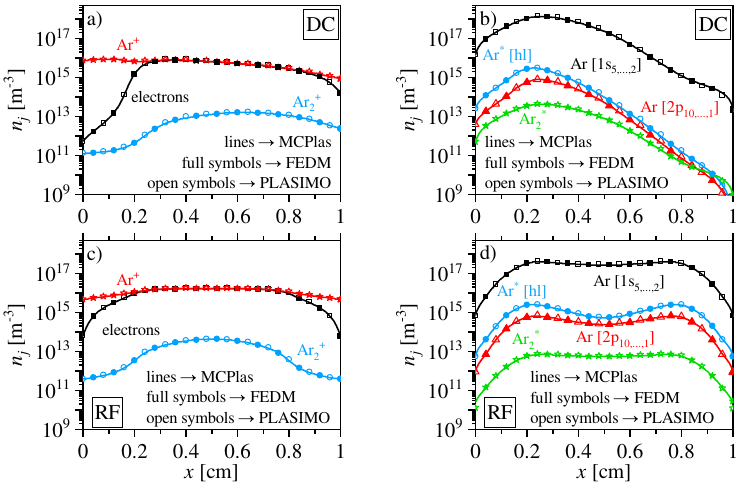}
	\caption{Comparison of modelling results obtained with MCPlas, PLASIMO and FEDM for the 23‑species RKM. The results show the stationary state of the DC (a, b) and the period-averaged state of the RF (c,d) glow discharges in argon for the density of the charged (a, c) and the neutral (b, d) species.
 	}
	\label{fig:figure_17}
\end{figure}

Modelling results for the DC and RF glow discharges in argon, obtained by MCPlas, PLASIMO, and FEDM with the 4-species argon RKM, are presented in figure~\ref{fig:figure_16}. The results for the DC case are presented for the steady state of the discharge. In the case of the RF discharge, the data are period-averaged in the stable periodic state of the discharge. The obtained values of particle species number densities and mean electron energy are in excellent agreement across all three modelling toolboxes. The same can be said for the 23-species RKM based on the results depicted in figure~\ref{fig:figure_17}. The largest difference between all modelling results obtained by MCPlas, PLASIMO and FEDM amounts to only few percentages, which proves the reusability of the proposed JSON RKM input data. It should be emphasised that this excellent agreement is achieved despite the fact that PLASIMO and FEDM are developed by different scientific groups and use different numerical methods. In particular, PLASIMO is based on the finite volume method and FEDM uses the finite element method. This comparison further demonstrates the reusability of the input data files.

\section{Conclusion}
\label{sec:conclusions}
MCPlas provides a transparent, fully traceable workflow for equation‑based fluid-Poisson plasma modelling in COMSOL. All information that defines a LTP model, i.e. reaction kinetics, species states and properties, transport and rate coefficients, boundary and surface parameters, and general plasma/source and diagnostics settings, is externalised into structured JSON documents that are validated against JSON schemas derived from the current LXCat and Plasma-MDS community developments. This systematic use of schema‑validated JSON input guarantees that COMSOL-model definitions are explicit, machine‑readable and interoperable, thereby aligned with the implementation of the FAIR data principles for plasma modelling data and enabling unambiguous reproduction of model setups.
This is achieved by automation of the construction of COMSOL models via LiveLink\textsuperscript{\texttrademark} \textit{for} MATLAB\textsuperscript{\fontsize{4}{4}\selectfont\textregistered}, giving full and transparent access to all equations and boundary conditions while preserving the convenience of a high‑level finite‑element environment. MCPlas supports multiple 1D and 2D geometries and several treatments of electron transport, particularly the novel drift-diffusion approximation (DDAn) not available in commercial packages. 

Code verification against the COMSOL Plasma Module using identical transport models and boundary conditions for DC and RF argon glow discharges showed excellent agreement, confirming the correctness of the generated models. When the enhanced electron transport description and advanced boundary conditions specific to MCPlas were enabled, substantial differences appeared in charged and excited species densities, mean electron energy and electric field distributions, particularly in cathod-sheath regions, demonstrating the physical impact of individual model formulations.
The capability of MCPlas to handle complex chemistries was demonstrated by switching, via JSON input only, from a simple 4‑species argon scheme to an extended 23‑species, 409‑process argon RKM, including individual 1s and 2p states and molecular species. This showed that detailed plasma composition and excitation dynamics can be accessed without manual re‑implementation of the model, thereby reducing errors and greatly facilitating verification and extension of published work. Finally, the same RKM input files in LXCat schema-based JSON format were used, without modification, in two independent plasma codes (PLASIMO and FEDM), yielding mutually consistent results for both DC and RF discharges. This cross‑platform consistency validates the interoperability and reusability of the standardized JSON input and confirms that MCPlas, together with its schema‑driven data structures, offers a robust, FAIR‑compliant foundation for reproducible, transparent and sharable plasma modelling workflows.

\section*{Data availability statement}
The data that support the findings of this study will be made publicly available after acceptance of the manuscript.

\section*{Code availability}
The source code of the MCPlas toolbox and all input data required to reproduce the presented examples are publicly available at the following URL/git repository: https://github.com/INP-SDT/MCPlas (git commit: 0b67eea)

\section*{Acknowledgment}
This work was partly funded by the Deutsche Forschungsgemeinschaft (DFG, German Research Foundation)—Project Numbers 213099267, 368502453, 407462159, 504701852, 509169873, 535827833.
 
\clearpage

%% References


\bibliographystyle{elsarticle-num}
\bibliography{reference}
%\the\textwidth \\


%% References


%\bibliographystyle{elsarticle-num}






% \end{thebibliography}

\end{document}